\documentclass[reprint,
superscriptadress,
showpacs, showkeys,floatfix, amsmath, amssymb,
aps,
jcp]{revtex4-1}

\usepackage{graphicx}
\usepackage[colorlinks,
    linkcolor={blue},
    citecolor={blue},
    urlcolor={blue}]{hyperref}

\usepackage{amsmath}
\usepackage{amssymb}
\usepackage{dcolumn}
\usepackage{bm}

\usepackage[english]{babel}
\usepackage{microtype}
\usepackage{indentfirst}
\usepackage{siunitx, booktabs}
\usepackage{tabularx}
\usepackage[caption=false]{subfig}
\usepackage{array}
\usepackage{setspace}
\usepackage{braket}
\usepackage{xcolor}
\usepackage{placeins}

\usepackage{amsfonts}
\usepackage{algorithmic}
\usepackage{textcomp}
\usepackage{xcolor}
\usepackage{xspace}
\usepackage{url}
\usepackage{listings}

\begin{document}

\title{Predicting the hydrogen bond strength from water reorientation dynamics at short timescales}%
\author{Frederik Zysk$^{1}$, Ana Vila Verde$^{2}$, Naveen K. Kaliannan$^{1}$, Kristof Karhan$^{1}$, and Thomas D. K\"uhne$^{3,4}$}
\email{t.kuehne@hzdr.de}
\affiliation{$^{1}$Dynamics of Condensed Matter, Chair of Theoretical Chemistry, University of Paderborn, 
Warburger Str. 100, D-33098 Paderborn, Germany
\\
$^{2}$Faculty of Physics, University of Duisburg-Essen, Lotharstrasse 1, 47057 Duisburg, Germany \\
$^{3}$Center for Advanced Systems Understanding (CASUS), Helmholtz-Zentrum Dresden-Rossendorf, 
Conrad-Schiedt-Str. 20, D-02826 Görlitz, Germany
\\$^{4}$Institute of Artificial Intelligence, Technische Universit\"at Dresden, Helmholtzstr. 10, D-01069 Dresden, Germany}%

\begin{abstract}
Path-integral molecular dynamics simulations and electronic structure-based energy decomposition analysis (EDA) are employed to connect hydrogen bond (H-bond) strength, its asymmetry, and the total delocalization energy at the water/air interface to experimentally measurable observables, such as the reorientation dynamics and the sum-frequency generation (SFG) spectrum. Using SFG spectra for distinct layers at the water/air interface, we validate the accuracy of our simulations and report a red-shift from the interface to bulk and a strongly bonded water peak at around 3250 cm$^{-1}$ in the layer closest to bulk. The reorientation dynamics of water molecules slow down from the interface to bulk, which correlates with the SFG results. From our EDA based on absolutely localized molecular orbitals, we observe a strong decline in total delocalization energy from bulk to the interface, as well as a decline in the strength of the strongest donor and acceptor interactions. The asymmetry between the two strongest interactions similarly rises towards the interface, while the importance of interactions from the outer solvation shells is greatly diminished and is lower than previously reported. 
Finally, we find that the strength of the strongest H-bond donor/acceptor is best correlated with the local minimum of the autocorrelation function resembling the L2 band librational motions. Following that, we propose a simple yet quantitative relationship between H-bond strength and the short-time reorientation dynamics at the water/air interface that could potentially be extended to predict H-bond strength in other hydrophobic systems from experimentally obtainable observables.      
\end{abstract}

\maketitle

\section{Introduction}
The water/vapor interface is ubiquitous in nature. It is present both in macroscopic length scales -- in oceans and freshwater resources -- and microscopic ones, e.g. in atmospheric~\cite{Andreae1997,Ghosal2005,Xu2008}, marine~\cite{Knipping2000,Ghosal2005}, and therapeutic~\cite{Haddrell2017} aerosols; in natural and artificial superhydrophobic surfaces~\cite{Jeevahan2018}, as well as in surfaces with the ability to adhere to hydrophobic surfaces in both wet and dry conditions~\cite{Hosoda2012}. Understanding the structure and dynamics of the  water/air interface is important to understand chemical and physical processes in  biology and aqueous chemistry~\cite{Ball, Bonn2015,Finlayson-Pitts2009}, as well as its role in promoting condensation reactions like peptide synthesis~\cite{deal2021water}.
The chemistry and physics of the air/water interface controls, e.g. the superhydrophobic nature of a surface coating~\cite{Jeevahan2018}, the production of gaseous Cl\textsubscript{2} in aqueous marine aerosols~\cite{Knipping2000}, and gas exchange~\cite{Liss1973,Merlivat1983}, to name just a few. These strong interfacial effects directly reflect the fact that interfacial water differs from water in the bulk.

A number of experimental and computational studies  have revealed that water molecules at the air/water, or air/aqueous solution interface have a different structure and kinetics than bulk water molecules~\cite{Kuhne2011,Jubb2012,Anna2012,Ni2013,Tong2013,Geissler2013,Venkateshwaran2014,Hsieh2011,Zhang2011a,Kessler,Bonn2015,pettersson2016,Naveen2020,MORITA2000,wilson2020kinetic,kappes2021chemistry,ojha2024nuclear}.
Most prior computational studies of the air/water interface have used simple empirical interaction potentials~\cite{morita2002theoretical,buch2005molecular,cipcigan2015hydrogen}. These potentials often perform far less well at other temperatures~\cite{Jorgensen1998}, and are typically unable to capture the change in polarization that water molecules experience at the interface. They capture nuclear quantum effects only implicitly, at the temperature and for the properties for which they were parameterized. Hence, studies using many-body molecular dynamics (MD)~\cite{dang1997molecular,medders2016dissecting}, path-integral MD (PIMD)~\cite{habershon2008comparison,Kessler,Spura} and semi-empirical methods~\cite{vysotsky2005quantum, john2016quantum} have risen in prominence. Both polarizability and nuclear quantum effects strongly contribute to the properties of water, including their temperature dependence~\cite{Kuo2006,Paesani2010,Paesani2011a,Spura,Ojha2018}, making ab-initio MD (AIMD) simulations including nuclear quantum effects highly desirable for the study of water~\cite{li2011quantum,spura2015fly,clark2019opposing}.  Because of the high computational cost  involved in simulating an interface, such direct simulations are a relatively new occurrence~\cite{Kessler,nagata2016surface}. 

The most salient feature of water is its ability to form hydrogen bonds (H-bonds)\cite{kühne2009static}. This ability is hindered when water meets an interface, with both simulation and experiment suggesting that approximately 20\% of water hydroxyl groups at the air/water interface are not H-bonded~\cite{Kuo2006,Hantal2010,Kuhne2011,Hsieh2011,Wang2011a,Anna2012,Tong2013,Kessler, ojha2025time}. Furthermore, X-ray absorption spectroscopy experiments and simulations suggest that the intermolecular O$\cdots$O distance increases by approximately 6\% at the air/water interface~\cite{Wilson2002a,Kuo2006}, which indicates that the strength of the remaining H-bonds in the interfacial region may differ from that in bulk or that interfacial water is mostly undercoordinated~\cite{Ishiyama2009,Bonn2015}.
The H-bond strength can be studied using information from the O-H stretching mode~\cite{Rey2002,Bakker2010,Perakis2016,ojha2018hydrogen}, as a weaker H-bond increases the O-H stretch frequency. 
Infrared (IR) absorbance and Raman scattering spectroscopy~\cite{Auer2007,Auer2008,Schmidt2007} are used to study H-bonds, but are not surface specific. The response will be noisy due to bulk molecules.
When studying only interfacial regions, various experimental studies use sum-frequency generation (SFG) spectroscopy as a convenient tool that combines IR and visible pulses to selectively probe interfacial molecules~\cite{Shen1989,Raymond2003,Gan2006,Levering2007,Bonn2015}. Symmetry selection rules lead to the cancelation of bulk responses. Computational MD simulations of the SFG spectra are used to gain insight into the structure and dynamics at the interface~\cite{MORITA2000,Ishiyama2009, Nagata2015,seki2019,ojha2019,tang2020,ojha2021hydrogen} and to validate the accuracy of water models~\cite{Liu2011,Medders2013,Ohto2019} reproducing experimental results. It has been found that the SFG response vanishes from a distance of 5~\AA~to the surface~\cite{Ishiyama2009} and that the property changes at the interface are based on an under-coordination of water molecules, leading to a "free" O-H peak at around 3700 $cm^{-1}$, and two H-bonded peaks at 3200 and 3400 $cm^{-1}$. Often, those two peaks are called "ice-like" and "liquid-like"~\cite{Du1994, Raymond2003,zhang2013}.

Directly characterizing the strength of H-bonds can now be done via energy-decomposition analysis methods such as those based on absolutely localized molecular orbitals (ALMO EDA)~\cite{khaliullin2013microscopic,kuhne2020cp2k}. This method decomposes interactions into physically meaningful components, enabling  deeper insights into inter-molecular bonding than the traditional total-energy electronic structure methods~\cite{Kitaura,Chen1996}. Previously, ALMO EDA has been used to investigate chemical bonding in molecular gas-phase complexes~\cite{Khaliullin2009, Ramos, Wang}. However, this method can also be used to investigate periodic condensed phase systems: specifically, it was used to quantify the contribution of the strongest donor-acceptor interactions to the average delocalization energy of a molecule~\cite{elgabarty2015covalency}. The two strongest donor and acceptor interactions are responsible for 93 \% of the total delocalization energy of a molecule, while there is a considerable asymmetry between those two strongest interactions~\cite{kühne2014nature,Elgabarty2020}.

As such, ALMO EDA provides invaluable insight into H-bond strength, though at a high computational cost; its output also cannot be directly compared with experiments. It is thus advantageous to look for observables that provide insight into H-bond strength and can be easily calculated from MD trajectories, and which are also experimentally accessible~\cite{ojha2018hydrogen}. Prior experimental and computational work has shown that the short time ($<$ 200~fs) reorientation dynamics of the water hydroxy groups (libration) are related to H-bond strength~\cite{Moilanen2008,Laage2006,Lawrence2003,Paesani2011a} and water heterogeneity~\cite{seki2019}. The librational motion is commonly described as forming a cone in the H-bond donor-acceptor direction. The semi-angle of the cone was reported to be inversely proportional to H-bond strength~\cite{Laage2006}.
  
In this work, we use PIMD simulations and density functional theory (DFT)-based ALMO EDA to connect H-bond strength, its asymmetry, and the total delocalization energy at the water/air interface to two experimentally measurable observables: The reorientation dynamics and the SFG spectrum. Using SFG spectra at the water/air interface, we validate the accuracy of our simulations. We investigate the reorientation dynamics of water molecules as a function of their position relative to the instantaneous air/water interface using simulations  that include nuclear quantum effects and relate those qualitatively to the SFG spectra. Moreover, ALMO EDA allows us to characterize the average H-bond strength, its asymmetry, and overall delocalization energy as a function of the position of each water molecule relative to the instantaneous air/water interface. Finally, we investigate the correlation between those properties and the short time ($<$ 200~fs) reorientation of water's hydroxy groups. In combination, we propose a simple quantitative relationship between H-bond strength and the short-time reorientation dynamics at the water/air interface that could potentially be extended to predict H-bond strength in other hydrophobic systems, taking the air/water interface as a benchmark~\cite{marx2004throwing}.

\section{Computational Methods}
\label{sec:computer}

\subsection{Path-integral molecular dynamics}
\label{sec:CD}

Our model of the water/vapor interface consisted of a bulk water part of $7\times7\times7$ molecules in a cubic cell with an edge length of 21.75 \AA. The cell was expanded in the z-direction by a factor of 5 to yield the final dimensions of $21.75\times21.75\times108.75$ \AA$^{3}$ to model the water/air interface.
The system is then equilibrated in the canonical NVT ensemble for 10~ps, followed by a centroid PIMD simulation of 8~ps in length. The simulation was performed at 300~K using the q-TIP4P/F model by Habershon et al.~\cite{Haberson2009}, including a three body correction (E3B) with 32 beads. For the E3B correction, code and parameters published by Tainter et al.~\cite{Tainter2014} were used. A discretized time step of 0.1~fs was chosen, whereas every fifth time step was written into the trajectory. The beads were contracted to 1 for both Ewald and Lennard-Jones interactions. All other interactions were computed using 32 beads. To average the data, a series of 250 statistically independent trajectories were computed. 

\subsection{Instantaneous water/air layers}
\label{layers}

To study the processes at the water/vapor interface, a reliable definition of the interface itself is needed. Using a surface definition based on time-averaged density functions neglects spatial fluctuations in space and time. The method of Willard and Chandler is employed here~\cite{Willard2010}. Instead of using a time-averaged density-field, a coarse-grained but time-dependent density-field in terms of Gaussian functions, located at the center of mass of the water molecules, is used. With this mechanism, it is possible to deduce the proximity of all water molecules from the instantaneous interface for every time step. The proximity $a_{i}$ of the $ith$ water molecule from the instantaneous interface for every step can be averaged over all instantaneous interfaces. The ensemble-averaged interface and the corresponding mean proximity are obtained as  
\begin{center}
\begin{equation}
\vec{a}_{i}={[\langle s \rangle -r_{i}]\cdot\langle n \rangle} \big\vert_{\langle s \rangle=\langle s \rangle_{i}}.
\end{equation}
\end{center}
Kessler et al.~\cite{Kessler} refer to the three distinct 3 \AA\ each as instantaneous water layers 0 to 2, from top (vapor) to bottom (bulk), which were deduced due to varying water configurations and orientations. Whereby 2.5 \AA\ of the topmost layer 0 are actually located in the vapor phase. While the molecules in layers 0 to 2 have a distinct structural order, phases beyond layer 2 do not obey any structural order and correspond to bulk water. Layer 0 cannot be viewed as a genuine water layer, but rather as a sparse population of water molecules with a higher proximity to the vapor phase than to the first water layer. In this study, however, too few water molecules resided in the corresponding layer at all times to make a reorientation dynamics analysis statistically viable. 

\subsection{SFG spectra calculations}
\label{sec:SFGcomp} 

The SFG spectra of the water/air interface reported in this work were calculated using the surface-specific velocity-velocity correlation function approach developed by Nagata and co-workers~\cite{Ohto2015, Naveen2020}. This approach is an alternative to the dipole moment-polarizability time-correlation function approach, which is computationally expensive and requires relatively long MD trajectories (several ns) to reach numerical convergence. Nevertheless, it has already been shown that the former formalism reproduces experimental SFG measurements for several aqueous interfaces~\cite{Naveen2020,Ohto2015,Ishiyama2009}. The equation used to calculate our SFG spectra is given by the following expression:
\begin{center}
\begin{equation}
\begin{aligned}
& X_{abc}^{res,(2)}(\omega)=
\frac{Q(\omega)}{i\omega^{2}} \\
&
\int_{0}^{\infty}dte^{-i\omega t}\times\left\langle\sum\limits_{i,j}g_{t}(r_{ij}(0);r_{t}) \dot{r}^{\text{OH}}_{c,i}(0)\frac{\dot\vec{r}_{j}^{OH}(t)\vec{r}^{\text{OH}}_j (t)}{|\vec{r}^{\text{OH}}_j(t)|}\right\rangle, \\
& if~a=b.
\end{aligned}
\end{equation}
\end{center}
\noindent where $r_{ij}(t)$ is the distance between the centers of mass of the O-H groups $i$ and $j$ at time $t$, and $g_{t} (r_{ij};r_{t})$ is the function that controls the cross-correlation terms with the cross-correlation cutoff radius $r_{t}$. If $a = b$ is not true, the entire term is set to zero. The intra-molecular distance and velocity of the O-H group $j$ at time $t$ are denoted as $\vec{r}_{j}^{OH}(t)$  and $\dot{\vec{r}_{j}}^{OH}(t)$ , respectively. The quantum correction factor Q(w) was taken from Ref.~\cite{Ohto2015} and the Hann window function was applied for the Fourier transformation of the time-correlation function. Even though non-Condon effects are neglected, all intra-molecular coupling effects are included~\cite{Naveen2020}. In our study, the cutoff $r_{t}$ is set to 2 \AA, thus including both auto- and intra-cross correlations of O-H modes. For every water layer, the 250 independent centroid PIMD trajectories were used to calculate the individual response, and the average of those 250 calculations is reported.

\subsection{Reorientation dynamics of O-H groups}
\label{sec:RDOH}

We calculate the second-order rotational dynamics using a second-order autocorrelation function similar to previous works by Vila Verde et al.~\cite{Anna2012} as
\begin{center}
\begin{equation}
P_{2}=\left\langle \frac{1}{2} (3\cdot \textrm{cos}^{2}(\vec{u}_{0}\cdot\vec{u}_{\tau})-1\right\rangle,
\end{equation}
\end{center}
where $\vec{u}$ is the vector characterizing the orientation of an O-H group. The maximum of the function is at 1, where a perfect orientation correlation occurs, and it has a minimum of -0.5, where all O-H groups are orthogonal to their initial orientation. $P_{2} = 0$ represents the state of perfect decorrelation. Only O-H groups at $t=0$ and $t=\tau$ are included. Vila Verde et al.~\cite{Ana2013} also used this method in a later study to investigate the reorientation dynamics of O-H groups in solutions of magnesium sulfate and cesium chloride. Vila Verde et al.~\cite{Anna2012} also found that a bi-exponential fit of the form 
\begin{center}
\begin{equation}
P_{2}(\tau)=a\cdot \exp\left(\frac{-t}{\tau_{s}}\right) + c\cdot \exp\left(\frac{-t}{\tau_{l}}\right)
\end{equation}
\end{center}
describes the reorientation dynamics reasonably well. The first term represents the short-time decay through librational motion, while the second term describes the long-time reorientation of O-H groups.

\subsection{Absolutely localized molecular orbitals energy-decomposition analysis}
\label{sec:newALMO}

The ALMO EDA method separates the total interaction energy of molecules $\Delta E_{tot}$ into
\begin{center}
\begin{equation}
\Delta E_{TOT}=\Delta E_{FRZ}+\Delta E_{POL}+\Delta E_{DEL}+\Delta E_{HO}, 
\end{equation}    
\end{center}
where E$_{TOT}$ is the total interaction energy of the unrelaxed electron densities on the molecules, and E$_{FRZ}$ is the orbital relaxation energy~\cite{khaliullin2013microscopic,kuhne2020cp2k}. Therein, E$_{POL}$ is the intramolecular relaxation associated with the polarization of the electron clouds on molecules in the field of each other, E$_{DEL}$ represents the two-body donor-acceptor orbital interactions, and E$_{HO}$ is a higher-order relaxation term. 
The two-body component
\begin{center}
\begin{equation}
\Delta E_{DEL}=\sum_{D,A=1}^{Mol}\Delta E_{D\rightarrow A}
\end{equation}
\end{center}
is the most interesting for our work. Each term in the summation arises from the delocalization of electrons from the occupied orbitals of donor D to the virtual orbital of acceptor A. These energies are obtained self-consistently and include cooperativity effects, which are the foundation of a correct description of the H-bond networks. As it is a natural descriptor of the H-bond network there is no need to employ arbitrary definitions for H-bonds based on geometry. 

To calculate the average H-bond energy in the various layers and the bulk, we used 3500 configurations from a 70~ps long DFT-based second-generation Car-Parrinello AIMD simulation with 384 water molecules in a $15.64\times15.64\times84.00$ \AA$^{3}$ slab~\cite{kuhne2007efficient, kühne2009static}. It was run in the canonical NVT ensemble with a step size of 0.5~fs. For each nuclear configuration, the five strongest acceptor and donor interactions of molecules were taken into consideration. For each layer, the average of the strongest and second strongest donor and acceptor interactions per water molecule is reported, as well as the average strength of the remaining interactions and the total average donor and acceptor interaction strength. Here, donor and acceptor are used to describe a molecule's role in the transfer of electron density. We also calculated the total average delocalization energy for water molecules in each layer and the asymmetry factors describing the difference in strength between the strongest and second strongest donor/acceptor interactions as
\begin{center}
\begin{equation}
\Upsilon_{D}=1-\frac{\Delta E_{D\rightarrow A^{2nd}}}{\Delta E_{D\rightarrow A^{1st}}}   
\end{equation}
\end{center}
for the donor asymmetry and 
\begin{center}
\begin{equation}
\Upsilon_{A}=1-\frac{\Delta E_{A\rightarrow D^{2nd}}}{\Delta E_{A\rightarrow D^{1st}}}  
\end{equation}
\end{center}
for the acceptor asymmetry, respectively~\cite{Kuhne2013}. Therein,
the first donor-acceptor (and also acceptor-donor) interaction is defined as the strongest interaction of each kind per water molecule. Values of $\Upsilon_{A}$ and $\Upsilon_{D}$ vary between 1, if there is only one donor or acceptor interaction per water molecule, and 0, if the two strongest interactions have the same value.

\section{Results}
\label{sec:res}

\subsection{Water layer SFG calculation}
\label{sec:SFGwater}

\begin{figure}
\begin{center}
\includegraphics[width=0.5\textwidth]{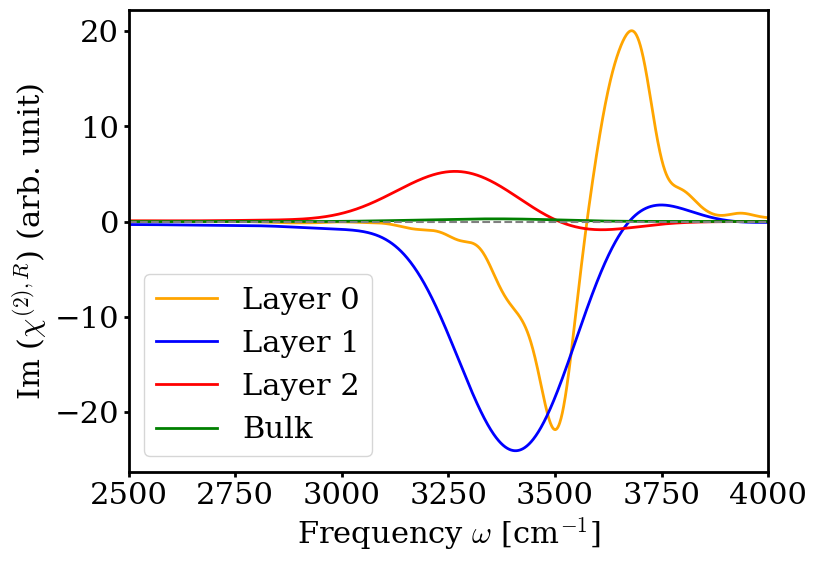}
\caption{\label{fig:SFG}SFG response for water layer 0 (orange), layer 1 (blue), layer 2 (red) and bulk water (green), respectively. The imaginary response is sensitive to the O-H group orientation. Negative values are indicates O-H stretches pointing towards the water bulk, whereas positive contributions are due to O-H stretches pointing towards the water vapor phase.}
\end{center}
\end{figure}

We can use SFG calculations (Figure~\ref{fig:SFG}) to validate our simulations and connect the features to the predominant water structures in the interface layers and to the H-bond strength directly~\cite{ojha2018hydrogen}. A positive imaginary response is based on O-H bonds pointing towards the surface, while a negative response is from orientations pointing away from the surface. Most newer experimental SFG studies of the water/air interface report a positive free O-H peak at around $3700~cm^{-1}$ and a negative H-bonded peak at around $3500~cm^{-1}$~\cite{Bonn2015}. Originally, another positive peak was reported at around $3200~cm^{-1}$~\cite{Ishiyama2009}, which was later attributed to being an artifact of the measurement~\cite{chiang2020,yamaguchi2015}. Overall, our results agree well with experimental and computational results~\cite{tang2020}. We divide the surface into layers, similarly to our previous work~\cite{Naveen2020}, to analyze the interface depending on its distance from the surface. 

Here, we can deduce how total SFG spectra arise from the different water layers of the interface. It is interesting to note that the most prominently discussed peaks at 3700 $cm^{-1}$ and 3500 $cm^{-1}$ stem from the topmost layer~\cite{Bonn2015,Ishiyama2009,chiang2020}, while the broadening of the $3500~cm^{-1}$ peak to $3400~cm^{-1}$ is due to layer 1. Those interactions are based on O-H stretches that are more strongly bonded than stretches in layer 0. We can still observe a small contribution to the "free" water peak at around 3700 $cm^{-1}$ from layer 1. We even see a positive response at around $3250~cm^{-1}$, close to previously reported peaks at around $3200~cm^{-1}$. The latter is solely located in layer 2, but it vanishes in the overall SFG response because the negative contribution from layer 1 is stronger. We conclude that we observe a red-shift in frequency from the surface (layer 0) to the bulk, which is based on the increasing strength of interactions. The bulk response vanishes, which confirms the estimation that water at a surface distance of 6.5~\AA has bulk properties that agree well with previous results showing bulk like properties at around 5.0~\AA~\cite{Ishiyama2009}.

\subsection{Reorientation Dynamics}
\label{sec:REOH}

\begin{table}
\caption{Bi-exponentially fitted short ($\tau_{s}$) and long ($\tau_{l}$) decay times for interface layers 1, 2 and bulk, respectively.}

{\begin{tabular}{lcr} \toprule  
 Layer \quad \quad & $\tau_{s}$[ps] \quad & \quad $\tau_{l}$[ps]\\ \midrule
1 & 0.407 & 7.750 \\
2 & 0.662 & 8.289 \\
Bulk & 0.622 & 8.937 \\ \bottomrule
\end{tabular}}
\label{table:bifitted}
\end{table}

Another observable that can be measured experimentally or calculated from simulations is the reorientation decay. With our computational approach, we calculate the dynamics in different layers and compare them to our SFG results from section~\ref{sec:SFGwater}. 
To analyze the reorientation dynamics of layers 1, 2, and bulk, we fit a bi-exponential function (see Section~\ref{sec:RDOH}) with $\tau_{s}$ and $\tau_{l}$ representing the short- and long-time dynamics. The corresponding values are shown in Table~\ref{table:bifitted}.

\begin{figure}
\begin{center}
\includegraphics[width=0.5\textwidth]{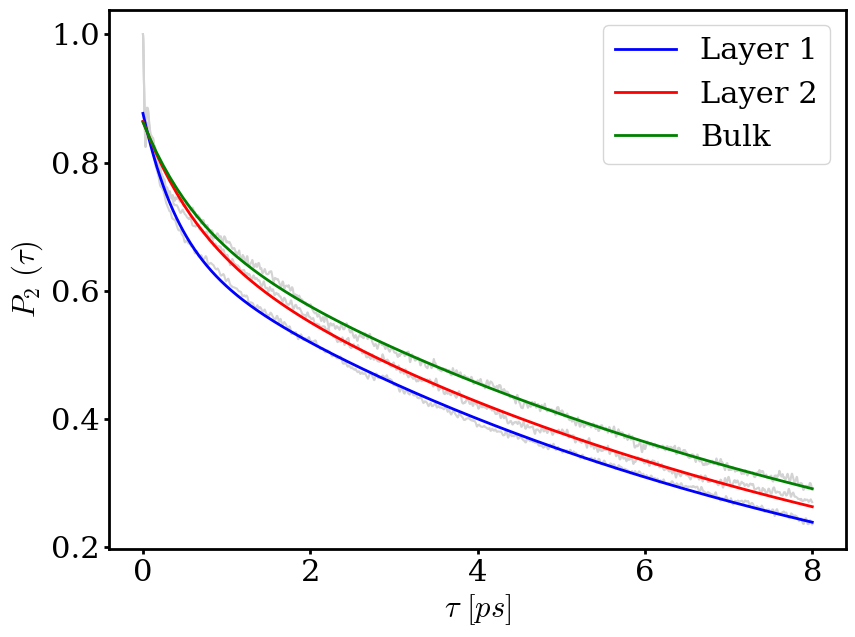}
\caption{\label{fig:Figure1}Reorientation dynamics of water hydroxy groups, as measured by the second order auto-correlation function $P_{2}(\tau)$ and a fitted bi-exponential function for layer 1 (blue), layer 2 (red) and bulk (green) at the water/air interface.}
\end{center}
\end{figure}

In Figure~\ref{fig:Figure1} an acceleration of long-time dynamics from bulk to the topmost interface layer can be seen. The long-time decay times $\tau_{l}$ in layer 2 and layer 1 are 7.3\% and 13.3\% faster than in bulk, respectively. Vila Verde et al. used a similar approach to study one distinct interface layer and bulk~\cite{Anna2012}. They reported 10\% (SPC/E) to 12\% (TIP4P) faster decay times at the interface than in bulk using classical MD simulations. Considering the splitting of their single interface layer into two distinct layers in our work, the results are qualitatively similar. The faster long-time reorientation decay times from the top layer to bulk agree qualitatively with the decrease in frequency reported in our SFG spectra for each layer in section~\ref{sec:SFGwater}. 

\begin{figure}
\begin{center}
\includegraphics[width=0.5\textwidth]{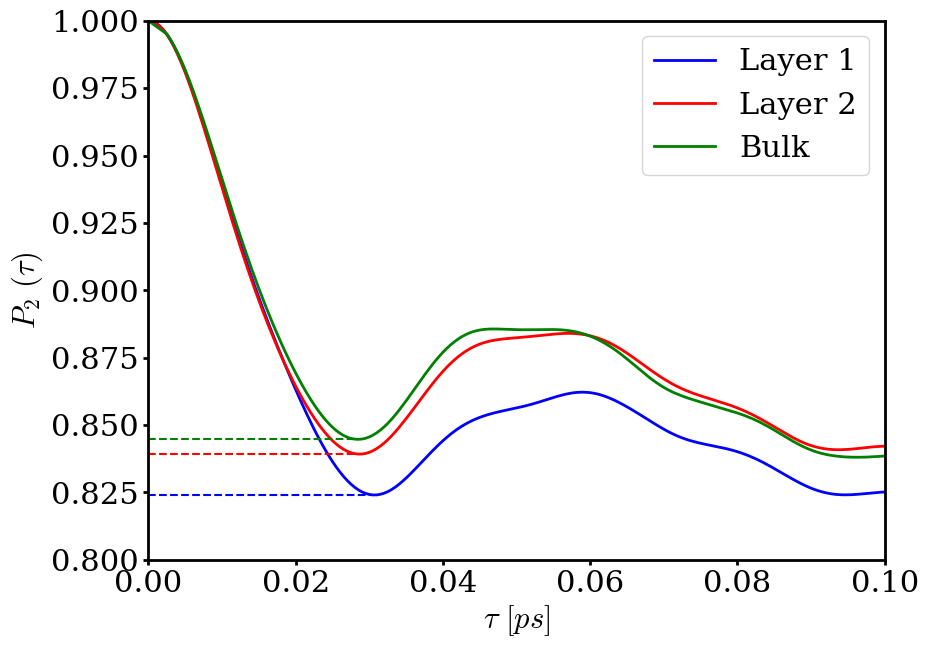}
\caption{\label{fig:Figure2}Short-time ($\le$ 0.1~ps) reorientation dynamics $P_{2}(\tau)$ for layer 1 (blue) and layer 2 (red) and bulk (green) at the water/air interface. Dashed lines mark the local minima labeled as $P_{2}(\tau_{L2})$.}
\end{center}
\end{figure}

The short-time dynamics for $\tau \le$ 0.1~ps can be seen in Figure~\ref{fig:Figure2}. There is a non-exponential decline at a similar rate for all layers and bulk, followed by under-damping.
The fitted short-time decay times are in the range of 0.41~ps to 0.66~ps (see Table~\ref{table:bifitted}), with the smallest value found in the topmost layer 1. Layer 2 and bulk exhibit 38.5\% and 34.5\% larger relaxation times, respectively, which equates to slower short-time reorientation dynamics. In this fit, layer 2 behaves very much like bulk in terms of short-time reorientation dynamics, but is actually slightly slower than in bulk, while layer 1 shows distinctly faster reorientation. This means the change in short-time reorientation decay does not qualitatively agree with our SFG results. This indicates that the short-time dynamics are either not strongly correlated with the SFG response or that the bi-exponential fit does not produce a representative observable. 

\subsection{Hydrogen bond energy and asymmetry}
\label{sec:H-Bond}

The previously discussed observables can either be obtained by theoretical or experimental means. However, the underlying source of both the reorientation dynamics and the SFG spectra is the H-bond network of water. This environment can be defined using ALMO EDA to calculate the interactions from the delocalization energies. 

\begin{table}
\caption{Average contributions of the strongest donor $E_{D_{1.}}$, second strongest donor $E_{D_{2.}}$ and subsequential donor interactions $E_{D_{3.- 5.}}$, as calculated from the five strongest donor interactions of each molecule. All energies are given in units of kJ/mol. The total donor interaction strength is also given as $E_{D_{all}}$ plus the corresponding asymmetry factor $\Upsilon_{D}$. The values are shown for each interface layer 0, 1, 2 and bulk water, respectively.}
{\begin{tabular}{lccccr} \toprule  
 Layer \quad & \quad $E_{D_{1.}}$ \quad \qquad & \quad $E_{D_{2.}}$ \quad & \quad $E_{D_{3.-5.}}$ \quad & \quad $E_{D_{all}}$ \qquad & \qquad $\Upsilon_{D}$\\ \midrule
0 & 13.687 & 1.109 & 0.164 & 14.960 & 0.919 \\
1 & 21.148 & 6.556 & 0.592 & 28.296 & 0.690 \\
2 & 23.357 & 9.928 & 0.974 & 34.259 & 0.575 \\
Bulk & 23.994 & 10.150 & 1.097 & 35.241 & 0.577\\ \bottomrule
\end{tabular}}
\label{table:ALMO1D}
\end{table}

In our ALMO EDA simulations, we have calculated the delocalization energies based on the five strongest donor and acceptor interactions of each molecule. In Table~\ref{table:ALMO1D}, the average donor interactions are shown. Donor interactions represent water molecules donating electron density to another water molecule. It can be seen that the overall donor interaction strength $E_{D_{all}}$ increases from interface layer 0 to bulk. As broken down in Table~\ref{table:ALMO1D}, this is the case for the strongest donor $E_{D_{1.}}$, the second strongest donor $E_{D_{2.}}$, and further interactions $E_{D_{3.-5.}}$ outside the first solvation shell.
This agrees nicely with our SFG calculations (see Figure~\ref{fig:SFG}), which show a red-shift of frequencies from the interface towards the bulk as donor interactions become stronger. 

The two strongest interactions are responsible for the vast majority of contributions to the overall delocalization energy, which is in accordance with earlier results~\cite{Kuhne2013}. In the case of water, further interactions contribute only around 3.1\% to the total delocalization energy in the bulk phase, 2.8\% in layer 2, 2.1\% in layer 1, and 1.1\% in layer 0, respectively. 
This is to say that the importance of $E_{D_{3.- 5.}}$ contributions decreases towards the interface.
Similarly, the increase of asymmetry towards the interface from 0.577 to 0.919 shows that the H-bond network is increasingly distorted when approaching the interface~\cite{Kuhne2011}.
This can also be seen in the SFG spectra, as bulk, which has the smallest asymmetry, shows no response, while layer 0, with the highest asymmetry, exhibits the most pronounced response (see Figure~\ref{fig:Figure1}).

This is further exemplified when comparing the strongest and second strongest interactions. The average strongest donor interactions in bulk are calculated to be 23.99~kJ/mol, which decreases towards the interface layer 0 to only 13.69~kJ/mol. This shows that the strongest donor interactions become weaker from bulk towards the interface. However, the second strongest interaction $E_{D_{2.}}$ weakens even more, going from 10.15~kJ/mol for bulk to only 1.11~kJ/mol for layer 0. 

\begin{table}
\caption{Average contributions of the strongest acceptor $E_{A_{1.}}$, second strongest interaction $E_{A_{2.}}$ and subsequential weaker interactions $E_{A_{3.-5.}}$, as calculated from the five strongest acceptor interactions of each molecule. All energies are given in units of kJ/mol. The total acceptor interaction strength is also given as $E_{A_{all}}$ plus the corresponding asymmetry factor $\Upsilon_{A}$. The values are shown for each interface layer 0, 1, 2 and bulk water, respectively.}
{\begin{tabular}{lccccr} \toprule  
Layer \quad & \quad $E_{A_{1.}}$ \quad \qquad & \quad $E_{A_{2.}}$ \qquad & \quad $E_{A_{3.- 5.}}$ \quad & \quad $E_{A_{all}}$ \qquad & \qquad $\Upsilon_{A}$\\ \midrule
0 & 12.448 & 0.573 & 0.150 & 13.171 & 0.954 \\
1 & 20.690 & 7.635 & 0.476 & 28.801 & 0.631 \\
2 & 22.701 & 10.669 & 0.669 & 34.041 & 0.530 \\
Bulk & 23.430 & 11.107 & 0.714 & 35.251 & 0.526\\ \bottomrule
\end{tabular}}
\label{table:ALMO1A}
\end{table}

We observe similar trends for the acceptor interactions shown in Table~\ref{table:ALMO1A}. It can be seen that the overall donor and acceptor delocalization energies are very similar for bulk (as should be the case), but they diverge further for the other layers. For layer 0, the total average donor contribution is around 1.8~kJ/mol larger than the acceptor contribution, while the donor contribution is around 0.5~kJ/mol per molecule larger in layer 1. This can be explained by the weaker acceptor interactions for layer 0 being donor interactions of molecules in layer 1. These interactions seem to be especially weak. As the number of molecules in layer 0 is lower than in layer 1, it can be concluded that more molecules in layer 1 act as weak donors to fewer molecules in layer 0. This interplay can be seen again in layer 2, which shows, on average, slightly weaker acceptor interactions that are based on the distortion seen in layer 1. 

\begin{table}
\caption{Values of the average total delocalization energy $E_{H}$ of a single water molecule, H-bond strength of the strongest donor interaction $E_{D_{1.}}$, autocorrelation function local minimum $P_{2}(\tau_{L2})$ for $\tau \le$ 0.05~ps and the long-time reorientation decay time $\tau_{l}$ in interface layers 0, 1, 2 and bulk, respectively. The energies $E_{H}$ and $E_{D_{1.}}$ are both given in units of kJ/mol.}
{\begin{tabular}{lcccc} \toprule
  & \multicolumn{2}{l}{ALMO EDA} & \multicolumn{2}{l}{Reorientation Dynamics}\\ 
   \midrule
 Layer \quad & \quad $E_{H}$ \qquad & \qquad $E_{D_{1.}}$ \qquad \qquad & \quad $P_{2}(\tau_{L2})$ \quad & $\tau_{l}$ \\ \midrule
 0 & 28.131 & 13.687 & -----  & -----\\
1 & 57.087 & 21.148 & 0.824 & 7.750 \\
2 & 68.300 & 23.357 & 0.839 & 8.289 \\
Bulk & 70.492 & 23.994  & 0.845 & 8.937\\ \bottomrule
\end{tabular}}
\label{table:ALMOcorr}
\end{table}

For further analysis, we also calculated the average total delocalization energy for a water molecule in its respective layer and report it in Table~\ref{table:ALMOcorr}. It is, similarly to previous results, highest in bulk at 70.49~kJ/mol, followed by 68.30~kJ/mol for layer 2, 57.09~kJ/mol for layer 1, and only 28.13~kJ/mol for the topmost layer 0, respectively. This gives an idea of the total strength of the bonding network in which a singular molecule is located. As such, water in each layer further away from the surface exists in a stronger water network, which we also showed by the red-shift in the SFG spectra. 

\subsection{Relationship between reorientation dynamics and hydrogen bond strength}
\label{sec:Relationship}

We extracted the SFG response, the short- and long-time reorientation dynamics, and various properties characterizing the interaction environment and H-bond strength. The main idea is to correlate the H-bond strength in different water layers with the short- and long-time reorientational dynamics. Here, the bi-exponentially fitted values for the short-time decay are outside the timeframe of the dynamics classified as librational motions. As the librational motions occur on timescales $\le$ 0.1~ps (see Figure~\ref{fig:Figure2}), fitted decay times of 0.4~ps to 0.6~ps are not representative of the actual phenomena. Short-time decay does not correlate with H-bond strength from ALMO EDA energies in Table~\ref{table:ALMOcorr}, with diverging trends for the water/air interface.

In fact, more useful information might be extracted by analyzing the unfitted data within that time-frame. Looking directly at the raw dynamics for times $\le$ 0.1~ps shown in Figure~\ref{fig:Figure2}, there is a steep loss of initial correlation until times $\le$ 0.03~ps. This can be seen as the movement of the O-H stretches in their librational cone. The value of that local minimum is based on the amplitude of this motion, as a larger decorrelation is equal to a larger amplitude/angle in the O–H stretch libration. 

Our aim is to study the correlation between this librational motion, which is an experimentally measurable observable of $P_{2}$, and the H-bond interaction strength quantified by our ALMO EDA simulations. The short-time decorrelation is extracted from Figure~\ref{fig:Figure2}, labeled as $P_{2}(\tau_{L2})$, and can be found in Table~\ref{table:ALMOcorr}. These values are correlated with our ALMO EDA energies in their corresponding layer and related properties, such as donor and acceptor asymmetry. Based on a simple least-squares fitting, we obtain a linear relationship between $P_{2}(\tau_{L2})$ and the average strongest donor interaction strength, i.e. 
\begin{center}
\begin{equation}
E_{D_{1.}} (kJ/mol)= 137.78\cdot P_{2}\left(\tau_{L2}\right) - 92.35. 
\end{equation}
\end{center}
The mean squared error (MSE) is calculated to be 0.007~kJ/mol, with a coefficient of determination (COD) of 0.996. Keep in mind that we are only fitting three data points here, but each data point was averaged over a large number of calculations, as we have described in Sections~\ref{sec:newALMO} and \ref{sec:CD}, respectively. Some other interesting properties to be evaluated are the total delocalization energy $E_{H}$ with a MSE of 0.562~kJ/mol and a COD of 0.984, the total donor delocalization energy $E_{D_{all}}$ with a MSE of 0.210~kJ/mol and a COD of 0.977, and the donor asymmetry $\Upsilon_{D}$ with a COD of 0.915. As all the values are properties describing interaction strength, they are decent predictors for the short-time dynamics, though the strongest donor interaction $E_{D_{1.}}$ was found to be the most predictive. 

We also study the correlation of the same properties with the long-time decay times $\tau_{l}$ shown in Table~\ref{table:ALMOcorr}, as obtained by a bi-exponential fitting.
Correlating the total delocalization energy $E_{H}$ with the long-time dynamics described by $\tau_{l}$, we obtain a MSE of 5.8~kJ/mol and a COD of only 0.83. Similarly, for the most predictive property $E_{D_{1.}}$ of the short-time dynamics, the correlation with $\tau_{l}$ is weaker, with a MSE of 0.186~kJ/mol and a COD of 0.87, respectively. Total donor strength and asymmetry in any given layer are even poorer predictors of long-time dynamics. This generally leaves the average strongest donor of a water molecule as the best predictor for the short-time and long-time reorientation dynamics. The short-time reorientation dynamics described by $P_{2}(\tau_{L2})$ are well predicted by energies from our ALMO EDA calculations and, especially, by the strongest H-bond donor, which is validated by SFG results that qualitatively agree with the changing dynamics.

\section{Conclusion}
\label{sec:conc}

We found that the total interaction energy of water molecules strongly decreases towards the surface, as does the H-bond strength of the strongest donor or acceptor interaction. Hence, water molecules exist in a more strongly bonded network the further away they are from the interface, while their H-bond asymmetry decreases. This was validated by our SFG calculations, which agree with experimentally and computationally produced spectra and show that layers with higher interaction energy exhibit red-shifted SFG responses. We also detected a strongly bonded water peak in layer 2 that is not visible when calculating the total SFG response. They also validate H-bond asymmetry calculations by showing that layers with higher asymmetry exhibit larger variances in SFG frequencies. The different layers also vary in their reorientation dynamics, slowing down for long-time dynamics towards the bulk. We found that the common method of using a bi-exponential fit of the reorientation dynamics does not accurately represent short timescales. To investigate the short-time reorientation dynamics, we found that the local minimum of the autocorrelation function resembling the L2 band librational motions is more descriptive of the short-time dynamics. We found a linear relationship between this new experimentally obtainable observable $P_{2}(\tau_{L2})$ and the on average strongest H-bond of a molecule in any given layer. The total delocalization energy of a given molecule is the second best predictor of short-time dynamics. Through this, we can predict the H-bond strength and total delocalization energy from experimental observables at hydrophobic interfaces, hopefully also extending to water in biological environments.  

\section*{Acknowledgement(s)}

The authors gratefully acknowledge funding by the Deutsche Forschungsgemeinschaft (DFG, German Research Foundation) under Germany's Excellence Strategy - EXC 2033 - 390677874 - RESOLV and the generous  allocation of computing  time  on  the supercomputer “Noctua” of the Paderborn Center for Parallel  Computing (PC2). 

\bibliography{ReorientationDynamics}

\end{document}